\documentstyle[twoside,fleqn,npb,epsfig]{article}

\newcommand{\beq}{\begin{equation}}
\newcommand{\eeq}{\end{equation}}
\newcommand{\beqa}{\begin{eqnarray}}
\newcommand{\eeqa}{\end{eqnarray}}
\newcommand{\s}{\hat s}
\newcommand{\th}{\hat t}
\renewcommand{\a}{\bar\alpha_s}

\newcommand{\AmS}{{\protect\the\textfont2
  A\kern-.1667em\lower.5ex\hbox{M}\kern-.125emS}}

\title{NLL BFKL and NNLO}

\author{Vittorio del Duca\address{Istituto Nazionale di Fisica Nucleare\\ 
Sezione di Torino\\ via P. Giuria, 1\\ 10125 - Torino, Italy}}

\begin{document}

\begin{abstract}
The BFKL resummation at LL and NLL accuracy is briefly reviewed,
with particular emphasis on the connection between the NLL corrections 
to the BFKL equation and exact NNLO calculations.
\end{abstract}

\maketitle

\section{LL BFKL}

In the limit of center-of-mass energy much greater than
the momentum transfer, $\s\gg|\th|$, any scattering process is dominated
by gluon exchange in the cross channel. Building upon this fact,
the BFKL theory~\cite{bal} models
strong-interaction processes with two large and disparate scales,
by resumming the radiative corrections to parton-parton
scattering. This is achieved
to leading logarithmic (LL) accuracy, in
$\ln(\s/|\th|)$, through the BFKL equation, i.e. a two-dimensional
integral equation which describes the evolution in transverse momentum
space and moment space of the gluon propagator exchanged in the cross
channel,
\beqa
&&\hspace{-7mm} \omega\, f_{\omega}(k_a,k_b)\, = \label{bfklb}\\
&&\hspace{-7mm} {1\over 2}\,\delta^2(k_a-k_b)\, +\, {\a\over \pi} 
\int {d^2k_{\perp}\over k_{\perp}^2}\, K(k_a,k_b,k)\, ,\nonumber
\eeqa
with $\a= N_c\alpha_s/\pi$, $N_c=3$ the number of colors, $k_a$ and $k_b$ 
the transverse 
momenta of the gluons at the ends of the propagator, and with kernel $K$,
\beqa
&&\hspace{-7mm} K(k_a,k_b,k) = \label{kern}\\
&&\hspace{-7mm} f_{\omega}(k_a+k,k_b) - {k_{a\perp}^2\over k_{\perp}^2 + 
(k_a+k)_{\perp}^2}\, f_{\omega}(k_a,k_b)\, ,\nonumber
\eeqa
where the first term accounts for the emission of a gluon of momentum
$k$ and the second for the virtual radiative corrections, which {\sl reggeize}
the gluon. Eq.~(\ref{bfklb}) has been derived in the
{\sl multi-Regge kinematics}, which presumes that the produced gluons are
strongly ordered in rapidity and have comparable transverse momenta.

The solution of eq.~(\ref{bfklb}), transformed from moment space to 
$y$ space, and averaged over the azimuthal angle
between $k_a$ and $k_b$, is
\beqa
&&\hspace{-7mm} f(k_a,k_b,y)\, = \int {d\omega\over 2\pi i}\, e^{\omega y}\, 
f_{\omega}(k_a,k_b)\label{solc}\\ &&= {1\over k_{a\perp}^2 } 
\int_{{1\over 2}-i\infty}^{{1\over 2}+i\infty} {d\gamma\over 2\pi i}\, 
e^{\omega(\gamma)y}\, \left(k_{a\perp}^2\over k_{b\perp}^2
\right)^{\gamma}\, ,\nonumber
\eeqa
with $\omega(\gamma)=\a\chi(\gamma)$ the leading eigenvalue of
the BFKL equation, determined through the implicit equation
\beq
\chi(\gamma) = 2\psi(1) - \psi(\gamma) - \psi(1-\gamma)\, .\label{llchi}
\eeq
In eq.~(\ref{solc}) the evolution parameter $y$ of the propagator is
$y=\ln(\hat s/\tau^2)$. The precise definition of the {\sl reggeization}
scale $\tau$
is immaterial to LL accuracy; the only requirement is that it is much
smaller than any of the $s$-type invariants, in order to guarantee that
$y\gg 1$.
The maximum of the leading eigenvalue, $\omega(1/2)=4\ln{2}\a$,
yields the known power-like growth of $f$ in energy~\cite{bal}.

What does the BFKL theory have to do with reality ? There is no evidence,
as yet, of the necessity of a BFKL resummation either in the scaling 
violations to the evolution of the 
$F_2$ structure function in DIS (for a summary of the theoretical status, see 
ref.~\cite{cat}), or in dijet production at large rapidity intervals \cite{mn}.
The most
promising BFKL footprint, as of now, seems to be forward jet production in DIS,
where the data~\cite{hera} show a better agreement with the BFKL resummation
\cite{hot} than with a NLO calculation \cite{mz}
(for a summary of dijet and forward jet production, see ref.~\cite{vdd}).

In a phenomenological analysis, 
the BFKL resummation is plagued by several deficiencies; 
the most relevant is that energy and longitudinal momentum
are not conserved, and since the momentum fractions of the incoming partons
are reconstructed from the kinematic variables of the outgoing partons,
the BFKL prediction for a production rate may be affected by large numerical
errors~\cite{ds}. However, energy-momentum conservation at each gluon
emission in the BFKL ladder can be achieved through a Monte Carlo 
implementation~\cite{bfklmc} of the BFKL equation~(\ref{bfklb}).

Besides, because of the strong rapidity ordering between the gluons 
emitted along the ladder, any two-parton invariant mass is large. Thus there
are no collinear divergences, no QCD coherence and no soft emissions
in the BFKL ladder.
Accordingly jets are determined only to leading order and have no 
non-trivial structure. Other resummations in the high-energy limit,
like the CCFM equation \cite{ccfm} which has QCD coherence and soft emissions, 
seem thus better suited to describe
more exclusive quantities, like multi-jet rates \cite{marche}.
However, it has been shown that, provided the jets are resolved, i.e.
their transverse energy is larger than a given resolution scale, the
BFKL and the CCFM multi-jet rates coincide to LL accuracy \cite{salam}.

\section{NLL BFKL and NNLO}

In addition to the problems mentioned above, the BFKL equation
is determined at a fixed value of $\alpha_s$ (as a consequence, the
solution (\ref{solc}) is scale invariant). All these problems can
be partly alleviated by computing the next-to-leading logarithmic (NLL) 
corrections to the BFKL equation~(\ref{bfklb}). In order to do that,
the real~\cite{real} and the one-loop~\cite{1loop,1loopeps,1loopds} 
corrections to the gluon emission in the kernel~(\ref{kern}) had to be 
computed, while the reggeization term in eq.~(\ref{kern}) needed
to be determined to NLL accuracy~\cite{2loop}. 

From the standpoint of a fixed-order
calculation, the NLL corrections \cite{real,1loop,1loopeps,1loopds,2loop} 
present features of 
both NLO and NNLO calculations. Namely, they contain
only the one-loop running of the coupling; on the other hand, in order to
extract the NLL reggeization term, an approximate evaluation of
two-loop parton-parton scattering amplitudes had to be performed \cite{2loop}.
In addition, the one-loop corrections to the gluon emission in the 
kernel~(\ref{kern}) had to be evaluated to higher order in the dimensional
regularization parameter $\epsilon$, in order to generate correctly all
the singular and finite contributions to the interference term between the
one-loop amplitude and its tree-level counterpart~\cite{1loopeps,1loopds}.
This turns out to be a general feature in the construction of the
infrared and collinear phase space of an exact NNLO calculation~\cite{bds},
and can be tackled in a partly model-independent way by using one-loop
eikonal and splitting functions evaluated to higher order in $\epsilon$
\cite{bdks}.

Building upon the NLL corrections \cite{real,1loop,1loopeps,1loopds,2loop} 
to the terms in
the kernel~(\ref{kern}), the BFKL equation was evaluated to NLL
accuracy~\cite{bfklnl,cc}. Applying the NLL kernel to the LL eigenfunctions,
$(k_\perp^2)^\gamma$, the solution has still the form of eq.~(\ref{solc}),
with leading eigenvalue,
\beqa
\omega(\gamma) &=& \a(\mu) \bigl[1-b_0\a(\mu)
\ln(k_{a\perp}^2/\mu^2) \bigr] \chi_0(\gamma) \nonumber\\ 
&+& \a^2(\mu) \chi_1(\gamma) \, \label{nllsol}
\eeqa
where $b_0=11/12 - n_f/(6N_c)$ is proportional the one-loop coefficient 
of the $\beta$ function, with $n_f$ active flavors, and $\mu$ is
the $\overline{MS}$ renormalization scale. $\chi_0(\gamma)$ is given in 
eq.~(\ref{solc}), and $\chi_1(\gamma)$ in ref.~\cite{bfklnl}. In 
eq.~(\ref{nllsol}),
the running coupling term, which breaks the scale invariance, has
been singled out.

Both the running-coupling and the scale-invariant
terms in eq.~(\ref{nllsol}) present problems that could undermine the
whole resummation program (for a summary of its status see ref.~\cite{carl}).
Firstly, the NLL corrections at $\gamma=1/2$ are negative and 
large~\cite{bfklnl} (however,  eq.~(\ref{nllsol}) no longer has a maximum 
at $\gamma=1/2$~\cite{ross}). Secondly, double transverse
logarithms of the type $\ln^2(k_{a\perp}^2/k_{b\perp}^2)$, which are not
included in the NLL resummation, can give a large contribution and need
to be resummed~\cite{doublelog}. Double transverse logarithms appear
because the NLL resummation is sensitive to the choice of reggeization
scale $\tau$; e.g. the choices $\tau^2=k_{a\perp}k_{b\perp}$, $k_{a\perp}^2$
or $k_{b\perp}^2$, which are all equivalent at LL, introduce double transverse
logarithms one with respect to the others at NLL. An alternative, but related,
approach is to introduce a cut-off $\Delta$ as the lower limit of
integration over the rapidity of the gluons emitted along the 
ladder~\cite{lip,delta}. This has the advantage of being similar in spirit
to the dependence of a fixed-order calculation on the factorization scale,
namely in a NLL resummation the dependence on the rapidity scale $\Delta$
is moved on to the NNLL terms~\cite{delta}, just like in a NLO exact 
calculation the
dependence on the factorization scale is moved on to the NNLO terms.

Finally, we remark that so far the activity has mostly been concentrated
on the NLL corrections to 
the Green's function for a gluon exchanged in the cross channel.
However, in a scattering amplitude this is convoluted with process-dependent
impact factors, which must be determined to the required accuracy.
In a NLL production rate, the impact factors must be computed at NLO.
For dijet production at large rapidity intervals, they are given in 
ref.~\cite{dsif}.

\end{document}